\documentclass[twocolumn,showpacs,preprintnumbers,amsmath,amssymb]{revtex4}

\usepackage{graphicx}
\usepackage{dcolumn}
\usepackage{bm}
\usepackage{url}

\begin{document}
\title{Fundamental Bounds and Approaches to Sequence Reconstruction \\ from Nanopore Sequencers}

\author{Jarek Duda}
\email{dudajar@gmail.com}
\altaffiliation{Institute of Computer Science, Faculty of Mathematics and Computer Science Jagiellonian University, Poland}
\author{Wojciech Szpankowski}%
 \email{spa@cs.purdue.edu}
\affiliation{Department of Computer Science and Center for Science of Information, Purdue University, USA}
\author{Ananth Grama}%
 \email{ayg@cs.purdue.edu}
\affiliation{Department of Computer Science and Center for Science of Information, Purdue University, USA}

\newcommand{\bea}{\begin{eqnarray}}
\newcommand{\eea}{\end{eqnarray}}
\newcommand{\be}{\begin{equation}}
\newcommand{\ee}{\end{equation}}
\newcommand{\beas}{\begin{eqnarray*}}
\newcommand{\eeas}{\end{eqnarray*}}
\newcommand{\bs}{\backslash}
\newcommand{\bc}{\begin{center}}
\newcommand{\ec}{\end{center}}
\def\SC {\mathscr{C}}

\begin{abstract}

Nanopore sequencers are emerging as promising new platforms for high-throughput
sequencing. As with other technologies, sequencer errors pose a major challenge
for their effective use. In this paper, we present a novel information theoretic
analysis of the impact of insertion-deletion (indel) errors in nanopore sequencers.
In particular, we consider the following problems: (i) for given indel error characteristics
and rate, what is the probability of accurate reconstruction as a function of sequence length;
(ii) what is the number of `typical' sequences within the distortion bound
induced by indel errors; (iii) using replicated extrusion (the process of passing a
DNA strand through the nanopore), what is the number of replicas needed to reduce the
distortion bound so that only one typical sequence exists within the distortion bound.

Our results provide a number of important insights: (i) the maximum length of a
sequence that can be accurately reconstructed in the presence of indel and substitution
errors is relatively small; (ii) the number
of typical sequences within the distortion bound is large; and (iii) replicated
extrusion is an effective technique for unique reconstruction. In particular, we
show that the number of replicas is a slow function (logarithmic) of sequence length --
implying that through replicated extrusion, we can sequence large reads using
nanopore sequencers. Our model considers indel and substitution errors separately. In
this sense, it can be viewed as providing (tight) bounds on reconstruction lengths
and repetitions for accurate reconstruction when the two error modes are considered in a
single model.
\end{abstract}
\maketitle


\section{Introduction}
\label{intro}

The past few years have seen significant advances in sequencing
technologies. Sequencing platforms from Illumina, Roche, PacBio and other vendors
are commonly available in laboratories. Accompanying these hardware advances,
significant progress has been made in statistical methods, algorithms,
and software for tasks ranging from base calling to complete assembly.
Among the key distinguishing features of these sequencing platforms are their
read lengths and error rates. Short read lengths pose problems for sequencing
high-repeat regions. Higher error rates, on the other hand, require oversampling
to either correct, or discard erroneous reads without adversely
impacting sequencing/ mapping quality. Significant research efforts have
studied tradeoffs of read-length, error rates, and sequencing complexity. An
excellent survey of these efforts is provided by Quail et al~\cite{sequencing_survey}.

More recently, nanopores have been proposed as platforms for sequencing.
Nanopores are fabricated either using organic channels (pore-forming proteins in
a bilayer) or solid-state material (silicon nitride or graphene). An ionic
current is passed through this nanopore by establishing an electrostatic
potential. When an analyte simultaneously passes through the nanopore, the
current flow is disrupted. This disruption of the current flow is monitored,
and used to characterize the analyte. This general principle can be used to
characterize nucleotides, small molecules, and proteins.
Complete solutions based on this technology are available from
Oxford Nanopore Technologies~\cite{e1}. In this platform, a
DNA strand is extruded through a protein channel in a membrane. The rate of
extrusion must be slower than current measurement (sampling) for characterizing each
base (or groups of small number of bases, up to four, in the nanopore at
any point of time).

In principle, nanopores have several attractive features -- long reads (beyond
100K bases) and minimal sample preparation. However, there are potential
challenges that must be overcome -- among them, the associated error rate.
The extrusion rate of a DNA strand through a protein channel is controlled
using an enzyme~\cite{Odonnell}. This rate is typically modeled as an
exponential distribution. When a number of identical bases pass through the
nanopore, the observed (non-varying) signal must be parsed to determine the
precise number of bases. This results in one of the dominant error modes for
nanopore sequencers. Specifically, insertion-deletion errors in such sequencers are
reported to be about 4\%~\cite{Odonnell,e4}. More recently, higher error rates
of approximately 12\% 
in addition to a 6\% substitution error have been reported~\cite{jain}.

The high error rate can be handled using replicated reads for de-novo
assembly, or through algorithmic techniques using reference genomes. The Oxford
Nanosequencer claims a scalable matrix of pores and associated sensors using
which replicated reads can be generated. Alternately, other technologies based on
bi-directional extrusion have been proposed. In either case, two fundamental
questions arise for de novo assembly: (i) for single reads, what is the bound
on read length that can be accurately reconstructed using a nanopore sequencer
with known error rates; and (ii) what is the number of replicas needed to
accurately reconstruct the sequence with high probability (analytically
defined).
In our analysis we assume that each fragment is read multiple times.
Since it is not currently possible to exercise such fine grain control over the
nanopore to read the same sequence multiple times, we can achieve this
through PCR amplification and resulting reads from multiple copies. These reads can be
aligned to achieve the same effect as reading fragments multiple times. Note
that the alignment problem is simpler here owing to longer
reads.

In this paper, we present a novel information theoretic analysis
of the impact of indel and substitution errors in nanopore sequencers. We model
the sequencer as a sticky insertion-deletion channel. The DNA sequence is fed into
this channel and the output of the channel is used to reconstruct the
input sequence. Using this model, we solve the following problems:
(i) for given error characteristics and rate, what is the
probability of accurate reconstruction as a function of sequence
length; (ii) what is the number of `typical' sequences within the
distortion bound induced by indels and substitutions; and (iii) what is the number of replicas
needed to reduce the distortion bound so that only one typical sequence exists
within the distortion bound (unique reconstruction).

Our results provide a number of important insights: (i) the maximum length of
sequence that can be accurately reconstructed in the presence of errors is
relatively small; (ii) the number of typical sequences within the distortion
bound induced by errors is large; and (iii) the number of replicas required for
unique reconstruction is a slow function (logarithmic) of the sequence
length -- implying that through replicated extrusion, we can sequence large
reads using nanopore sequencers. The bounds we derive are fundamental in
nature -- i.e., they hold for any re-sequencing/ processing technique.



\section{Approach}

In this section, we present our model and the underlying concepts in information theory
that provide the modeling substrates. We define notions of a channel, reconstruction,
an insertion-deletion channel, and distortion bound. We then describe how these concepts
are mapped to the problem of sequence reconstruction in nanopore sequencers.

Our basic model for a nanopore sequencer is illustrated in Figure~1. A
DNA sequence is input to the nanopore sequencer. This sequence is read and suitably
processed to produce an output sequence. We view the input sequence as a sequence of
blocks. Each block is comprised of a variable number ($k$) of identical bases. The
nanopore sequencer potentially introduces errors into each block by altering the number
of repeated bases. If the output block size $k'$ is not the same as the input block size
$k$, an indel error occurs. Specifically, $k' < k$ corresponds to a deletion error, and
$k' > k$ to an insertion error. Please note that this model does not account for
substitution errors.

\begin{figure*}[t!]
        \label{model}
    \centering
        \includegraphics[scale=0.6]{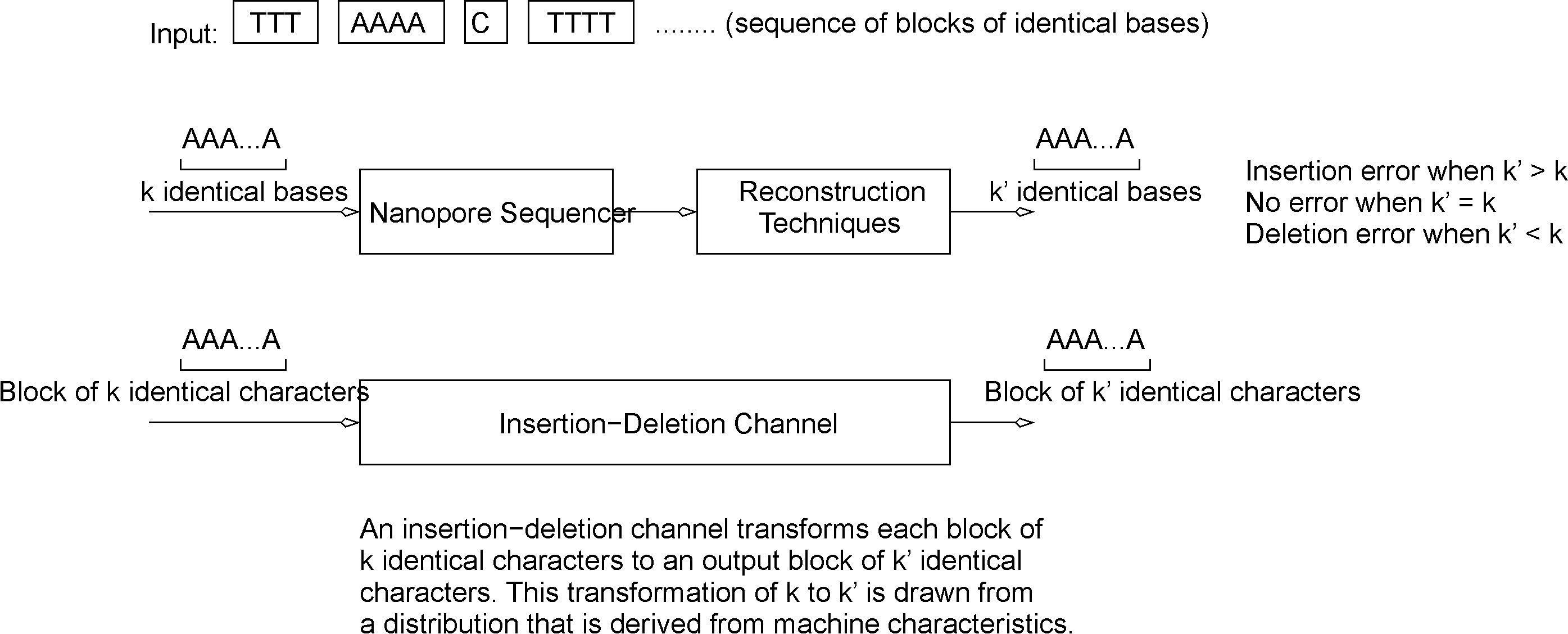}\
\begin{center}
        \caption{Overview of the proposed channel and its correspondence with a sequencer.}
\end{center}
\end{figure*}

We model the sequencing process (both the sequencer and the associated processing) as
a channel. A channel in information theory is a model (traditionally for a storage or
communication device, but in our case, used more generally) for information transfer with
certain error characteristics. The input sequence of blocks is sent into this channel.
The error characteristics of the channel transform a block of $k>0$ characters into a
block of $k'$ characters. This transformation is modeled as a distribution: $k' = G(k, P)$,
where $G$ is the distribution and $P$, the associated set of parameters tuned to the
sequencing platform. In typical scenarios,
the distribution peaks at $k$ and decays rapidly on either side. The distribution may
be asymmetric around $k$ depending on relative frequency of insertion and deletion
errors. We refer to such a channel as a sticky channel if it maintains the structure of blocks:
$k'>0$.

\paragraph{Insertion-Deletion Channels} In ideal communication systems, one often assumes
that senders and receivers are perfectly synchronized -- i.e., each sent bit is read
by the receiver. However, in real systems, such perfect synchronization is often not
possible. This leads to sent bits missed by the receiver (a deletion error), or read more
than once (an insertion error). Such communication systems are traditionally modeled
as insertion-deletion channels. Formally, an independent insertion channel is one in
which a single bit transmission is accompanied with the insertion of a random bit
with a probability $p$. An independent deletion channel is one in which a transmitted
bit can be deleted (omitted from the output stream) with a probability $p'$. An
insertion-deletion channel contains both insertions and
deletions~\cite{mitz}. Please note that a number of basic characteristics
of insertion-deletion channels, such as their capacity, are as-yet unknown in
information theory literature as well.

We consider a variant of the independent insertion-deletion model that is better
suited to nanopore sequencers. In particular, we recognize the primary source of
error in nanopore sequencers is associated with disambiguating the exact number of
identical bases passing through the nanopore. We modify the independent insertion-deletion
channel to the sticky insertion-deletion channel described above
(Figure~1). Coincidentally, the analysis of this block modification
insertion-deletion channel is easier -- as we demonstrate in this paper.

\paragraph{Typical Sequences.} There are $4^n$ distinct nucleotide sequences of $n$ bases,
each generated with a corresponding (idealized) probability of $4^{-n}$. For convenience, we
can also view this as probability as $2^{-2n}$, with the understanding that if each base is equally
likely, we would need two bits for each base. However, from asymptotic equipartition
property (AEP), we know that there is a {\em typical set} such that the probability of generating
a sequence belonging to this set approaches 1. In other words, while there may be sequences
outside of this set whose individual probability may be high, their number is small enough
that the total probability is dominated by the sequences in this typical set. Furthermore,
we know that the probability of drawing a sequence of length $n$ from this set is asymptotically given by
$2^{-H(X)n}$, where $H(X)$ is the entropy of the source. Comparing this expression with
the sequence probability assuming each base is equally likely ($2^{-2n}$), we note the
unsurprising conclusion that for our idealized case $H(X) = 2$. However, a number of studies
have shown that the entropy of living DNA is in fact much lower, as low as 1.7 or
below~\cite{Lanctot:2000:EDS:338219.338586}. This suggests that the number of typical
sequences is in fact much less than the number of total sequences. We use this notion of
{\rm typical sequences} and the {\em typical set} in our derivation of
the performance bounds.

\paragraph{Distortion Bound and Unique Reconstruction.} Viewing a DNA sequence passing
through a nanopore as a point (in some very high dimensional space), passing it through
a sequencer introduces an error. This error can be viewed as a hypersphere around the
original sequence. Each point in the hypersphere corresponds to a possible input sequence
with a probability that can be analytically quantified. We refer to this hypersphere as
a distortion ball. Ideally, we want the radius of this distortion ball to be as small
as possible -- containing only a single point. A weaker condition is that the distortion
ball contains only a single typical sequence. We refer to the former as an accurate
reconstruction and the latter as a unique reconstruction.

A single pass through the nanopore induces a distortion ball whose probability profile
can be quantified. We show that the radius of this ball can be reduced by replicated
extrusion through the nanopore. One of the key contributions of this paper is that
the radius shrinks rapidly with the number of extrusions, thus enabling accurate and/or
unique reconstruction with relatively small number of replicas.

\section{Methods}

\subsection{Notation and Theoretical Model} \label{notation}

The input sequence to the channel/ sequencer is drawn from an alphabet
$\mathcal{A}$. For DNA sequencing, $\mathcal{A} = \{A, T, C, G\}$.
The alphabet size $|\mathcal{A}|$ is denoted by $m$.
We assume that an $n$ length input
sequence $X$ is independent and identically distributed (i.i.d.); i.e.,
each sequence has the same probability distribution as the others and
all sequences are mutually independent. Mathematically, probability
$\Pr(x_i=s)=:p_s,\ \sum_s p_s=1$.

\subsubsection{Blocking Identical Symbols}
\label{block}

As mentioned, we view input and output sequences as sequences of blocks, with
each block comprised of one or more identical symbols. If $s^k$ denotes
$k\geq 1$ repeats of symbol $s$, we can write sequence $X$ as concatenation
of $N\leq n$ blocks: $X=s_1^{k_1} \ldots s_N^{k_N}$, such that $k_i>0$,
$s_{i+1}\neq s_i$. For example, a sequence $X=$ ``AATATTAA" is represented in
the block form as $A^2TAT^2A^2$.

We initiate our discussion by enumerating basic statistical properties
of block sequences. We see that for $s\neq s'$:
\be \Pr(s_{i+1}=s'|s_i=s) =\frac{p_{s'}}{1-p_s}. \ee

Since we have a block of symbols $s$, the block must start with
at least one appearance of symbol $s$. The probability that the block will have
length $k\geq 1$ is given by:
\be P_{sk}:= \Pr(k_i=k|s_i=s)=p_s^{k-1} (1-p_s)  \ee
\be \textrm{satisfying}\qquad \sum_{k\geq 1} P_{sk}=1. \nonumber \ee
The expected length of a block of symbols $s$ is given by:
$$\overline{k}_s:=\sum_{k\geq 1} kP_{sk} = \sum_{k\geq 0} k p_s^{k-1} (1-p_s)=\frac{1}{1-p_s}.$$
The expected number of symbols $s$ in the entire sequence of length $n$ is $n p_s$.
Therefore, the expected number of blocks of type $s$, and of all blocks is, respectively,
\be N_s:=n p_s/\overline{k}_s=np_s(1-p_s),\ee \be N:=\sum_s N_s=n\left(1-\sum_s p_s^2\right). \ee

\subsection{Sticky Insertion-Deletion Channel (indel)}

We model the sequencer using a sticky insertion-deletion channel. In this
channel, a block of $k$ consecutive identical symbols,
with probability $q_{kl}$, is transformed into a block of $l$ copies of the symbol:
\be \Pr(s^l|s^k)=q_{kl}(s)\qquad
 \textrm{where}\quad\sum_l q_{kl}(s)=1. \ee
The term sticky is used to imply that the block structure remains
unchanged; i.e., $q_{k0}(s)=q_{0l}(s)=0$. \\
In this model, $\{q_{kl}(s)\}$ specifies the block length change probabilities.
We can assume that for given $k$, $q_{kl}(s)$ has a maximum at $l=k$; i.e., the
probability that there is no error in a block exceeds any other (erroneous)
transformation. In Section \ref{experiment}, we consider two specific choices for
function $q$: an exponential distribution and independent insertion-deletions.
In this model, the probability of observing
an output sequence $Y$ from our indel channel is given by:
$$ \Pr(Y=s_1^{l_1}s_2^{l_2}...s_N^{l_N}|X=s_1^{k_1}s_2^{k_2}...s_N^{k_N})=\prod_{i=1}^N q_{k_i l_i}(s_i).$$
For example, true sequence $X=$ ``AAAATTAAA" $ = A^4 T^2 A^3$ is read by the sequencer
as $Y=$ "AATTTAA" $=A^2 T^3 A^2$ with probability $q_{42}(A)\cdot q_{23}(T)\cdot q_{31}(A)$.\\
For simplicity, we assume that $q$ is identical for different symbols
$s$: \be q_{kl}(s)\equiv q_{kl}.\ee
This implies that insertion-deletion errors in sequencing are independent of
the bases. Please note that this assumption is not a limitation of our framework.
Assuming $q_{k0} = q_{0l} = 0$, the number of blocks and their order remain
the same while applying the channel. However, the lengths of blocks may change.
The sequencing problem can then be reduced to the following task: determine the original
set of block counts $(k_i)_{i=1..N}$ from the following two cases under consideration:
  (i) the single extrusion case -- a single read sequence $Y$: $(l_i)_{i=1..N}$; and
  (ii) the multiple extrusion case -- each sequence is replicated $c\in \mathbb{N}$ times:
$(l^j_i)_{i=1..N}$ for $j=1,\ldots c$.

\subsection{Accurate Reconstruction from Nanopore Sequencers}
\label{mostprob}

We now consider the problem of constructing a true sequence (from the channel or the
sequencer) from an observed sequence (from the ionic current measurement).
This problem is one of reconstructing a true block sequence from an observed
block sequence at the output of the sticky channel. Since we assume that the block structure
is not changed by the channel, this problem can be solved one block at a
time. Specifically, we observe a block at the output of the channel of length
$l$ and we must infer the block length of the corresponding input, $k$.
We refer to this as the problem of finding the most probable reconstruction.

\subsubsection{Single Run Estimation: Inferring $k$ from a Single $l$}

In the first instance of the problem, we do not consider any replicas -- i.e., a
single block passes through the channel (sequencer) only once. From this single
observation of $l$, we must infer $k$. Assuming $n\to \infty$, the probability that an output block of
length $l$ was observed from an input block of length $k$ is given by:
\be Pr(s^k|s^l)=\frac{P_{sk}q_{kl}}{\sum_{k'} P_{sk'}q_{k'l}}=\frac{p_s^{k-1}q_{kl}}{\sum_{k'}p_s^{k'-1}q_{k'l}} \label{e1}\ee
where $P_{sk}=\Pr(k_i=k|s_i=s)=p_s^{k-1} (1-p_s)$.

In this case, the most likely input block length $k$ for observed output
block length $l$ is the one that maximizes $p_s^{k-1} q_{kl}$ for given $l$.
Let us denote this $k$ by $k_l$. We then have:
\be p_s^{k_l-1} q_{k_l l} = \max_k p_s^{k-1} q_{kl} \ee
We would expect that $k_l=l$. However, for a general distribution $q$, this is not
necessarily the case. The natural condition for $q$, given by $\max_k q_{kl} =q_{ll}$,
turns out not to be sufficient for this purpose. Even with this condition,
a single input block length $k$ may correspond to multiple corresponding observed block lengths $l$
(for some $l\neq l'$, $k_l=k_{l'}$)
, and some input block lengths $k$ might not have corresponding values of $l$ at all ($\forall_l k_l\neq k$).
Consequently, we need a stronger condition. It is easy to see that:
\be \forall_{l,i,s}\  q_{l-i,l}<(p_s)^i\cdot q_{l,l} \quad \Rightarrow \quad k_l = l \label{cond}\ee

We can now determine the probability that an observed block is properly corrected as:
$$m_k:=\sum_{l:k_l=k} q_{kl}\qquad (\textrm{0 if $k$ cannot be obtained}),$$
which reduces to $m_k=q_{kk}$ if (\ref{cond}) is satisfied. The expected number of
blocks of symbol $s$ is $N_s=n p_s (1-p_s)$. Their expected total length is $np_s$.
Therefore, the probability that we accurately correct all blocks is asymptotically given by:
\be \prod_s \left(\sum_k P_{sk} m_k\right)^{N_s}=2^{n\sum_s p_s(1-p_s)\lg(\sum_k P_{sk}m_k)}
\label{e2}\ee It is easy to see that this probability decreases exponentially with the
length of the sequence. Stated otherwise, this result shows that the probability that
we accurately reconstruct the entire sequence decreases exponentially in the
length of the sequence.

\subsubsection{Multiple runs: estimating $k$ from multiple $l$}

We now investigate how reading the same block multiple times can help infer the input
block length. We assume that each block is read $c$ times. This can
be done by extruding the same sequence through an array of nanopores.
In this case, we have $c$ observed values of $l_i$ (the $i$th block length),
$(l^j_i)_{i=1..N}$ for $j=1,\ldots c$. The probability that the corresponding
input block length is $k$ is given by:
\be Pr(s^k|s^{l^1},..,s^{l^c})=\frac{P_{sk}q_{kl^1}...q_{kl^c}}{\sum_{k'} P_{sk'}q_{k'l^1}...q_{k'l^c}}.\ee
Let us analogously define $k_{\mathbf{l}}$, where $\mathbf{l}:=\{l^1,..,l^c\}$, as the input block length $k$ that maximizes
$p_s^{k-1}\cdot q_{kl^1}...q_{kl^c}$, given by:
$$ p_s^{k_{\mathbf{l}}-1} q_{k_{\mathbf{l}} l^1}...q_{k_{\mathbf{l}}l^c} = \max_k p_s^{k-1} q_{kl^1}...q_{kl^c}=
 \max_k p_s^{k-1} \prod_i q_{ki}^{\tilde{l}^i}$$
\be = \exp\left(\max_k \ (k-1)\ln(p_s)+\sum_i \tilde{l}^i \ln(q_{ki})\right) \label{cross}\ee
where $\tilde{l}^i=\#\{j:l^j=i\}$ is empirical probability distribution.
The $\sum_i \tilde{l}^i \ln(q_{ki})$ term is equal to minus cross entropy between observed frequency $\{\tilde{l}^i\}_i$ and probability distribution expected for a given $k$: $\{q_{ki}\}_i$.
The $(k-1)\ln(p_s)$ term favors shorter blocks.
For an observed empirical distribution $\{\tilde{l}^i\}_i$, we should choose $k$ corresponding
to the closest probability distribution $\{q_{ki}\}_i$, as the one
maximizing $(k-1)\ln(p_s)+\sum_i \tilde{l}^i \ln(q_{ki})$.

The probability that input block length $k$ is accurately determined is given by:
\be m_{kc}:=\sum_{\mathbf{l}:k_{\mathbf{l}}=k} q_{kl^1}..q_{kl^c}.\ee
As before, the probability that we accurately infer all block sizes (accurate sequencing)
in analogy to (\ref{e2}) decreases exponentially as:
\be 2^{n\sum_s p_s(1-p_s)\lg(\sum_k P_{sk}m_{kc})}. \ee
Unfortunately the problem of finding $k_{\mathbf{l}}$ is a complex estimation procedure,
and therefore finding the required number of replicas, $c$, for unique reconstruction
appears difficult. However, as we show next, using tools from information theory, we
can estimate this replication rate. More importantly, we show that this replication rate is a slow
function of sequence length.


\subsection{Fundamental Bounds for Unique Reconstruction}
\label{inftheory}

We rely on an information theoretic approach to computing the minimum number of replicas $c$
required for accurate reconstruction. We do this by modifying the original problem somewhat.
Recall that the noise model of the channel introduces a distortion ball around the output
sequence. It is possible that multiple input sequences belong in this distortion ball --
leading to the problem of identification of the most probable reconstruction. However,
if we could use replicated extrusion of blocks to shrink the distortion radius to the point
where only one sequence belongs in the ball, we have unique reconstruction. We use this
principle to focus on the problem of number of {\em typical sequences} that belong in the
distortion ball, and find the number of replicas $c$ for which this number approaches one.
Please note that this problem is slightly distinct from the problem of most probable
reconstruction.

\subsubsection{Difference Between the Most Probable and Typical Reconstruction}

We begin our discussion by highlighting the differences between the {\em most probable}
and {\em typical} reconstructions. To gain an intuition about the difference between
these two types of reconstructions, let us briefly look at error correction
of the basic binary symmetric channel (BSC): we send $N$ bits, each of them
has independent probability $\epsilon<1/2$ of being flipped. Observe that we
could write this in the formalism we have introduced for blocks as:
$k,l\in\{0,1\},\ q_{00}=q_{11}=1-\epsilon,\ q_{01}=q_{10}=\epsilon$.

Obtaining from the output sequence $Y\in \{0,1\}^N$, the most probable input
sequence $X$ is simple -- it is simply $X = Y$. The probability that this input
sequence $X$ is the correct sequence is given by: $(1-\epsilon)^N=2^{N\lg(1-\epsilon)}$.
However, for large $N$, we expect that approximately $\epsilon N$ bits are flipped. There
are
$${N \choose \epsilon N}\approx 2^{Nh(\epsilon)} $$
$$ \left(\textrm{where}\quad h(p)=-p \lg(p)-(1-p)\lg(1-p)\right)$$
different ways of doing it. These are all {\em typical corrections} - which are
statistically dominating for large $N$ (Asymptotic Equipartition Property). In this case,
it is easy to see that conditional entropy $H(X|Y)=N \cdot h(\epsilon)$.
Having no additional information, all of these typical corrections are equally probable.
Consequently, the probability of choosing the correct one is given by
$2^{-H(X|Y)}=2^{-Nh(\epsilon)}$. Conversely, the definition of the channel says
that the probability that a given typical correction (flipped $\epsilon N$ bits)
is the correct one is given by:
$$\epsilon^{N\epsilon}\cdot (1-\epsilon)^{N(1-\epsilon)}=2^{-N h(\epsilon)}=2^{-H(X|Y)}$$
 -- exactly the same as before.

To summarize, typical corrections correspond to a Hamming sphere $S(Y,\epsilon N)$. The
most probable correction corresponds to its center and for unique reconstruction,
this sphere should reduce to a point; i.e., $H(X|Y)\approx 0$.

\subsubsection{Entropy in the Framework of Blocks of Identical Symbols}

Returning to our original problem, by definition of a typical sequence, the number of
typical sequences of length $n$, $X^n$, grows asymptotically as $\exp(H(X^n))$, where
\be H(X^n)=nh^x\qquad \textrm{and}\qquad h^x:=-\sum_s p_s\ln(p_s).\label{entr}\ee

We now derive this entropy formula in the block framework for the asymptotic case
(large $n$). This analysis is analogous to the result of Mitzenmacher et al.
\cite{mitz}, where the non-asymptotic case is presented. We must deal with two
additional considerations here: our alphabet is not binary; and the distribution
among input symbols is not necessarily uniform.

The information contained in the input sequence in the block framework:
$X^n= s_1^{k_1}\ldots s_N^{k_N}$ can be split into two parts -- the sequence of
symbols ($s_i$), and corresponding block lengths ($k_i$). $H(X^n)$ is sum of the two
entropies. Using results from Section (\ref{block}), the entropy of selecting the symbol
for succeeding block ($s_{i+1}\neq s_i$) is given by:
$$ h_s \equiv H(s_{i+1}|s_i=s)=-\sum_{s'\neq s} \frac{p_{s'}}{1-p_s}\ln\left(\frac{p_{s'}}{1-p_s}\right) $$
$$ =\frac{1}{1-p_s} \sum_{s'\neq s} p_{s'}(\ln(1-p_s)-\ln(p_{s'}))$$
$$ =\frac{1}{1-p_s}\left((1-p_s)\ln(1-p_s)+h^x+p_s\ln(p_s) \right) $$
\be =\frac{h^x-h(p_s)}{1-p_s}.\ee
The entropy of choosing block length for symbol $s$ is given by:
$$h^x_s\equiv H(k_i|s_i=s)=-\sum_{k\geq 1} P_{sk} \ln(P_{sk}) $$
$$ = -\sum_{k\geq 1} p_s^{k-1} (1-p_s) \ln\left(p_s^{k-1} (1-p_s)\right)$$
\be =-\frac{p_s}{1-p_s}\ln(p_s)-\ln(1-p_s)
 =\frac{h(p_s)}{1-p_s}.\ee
because $\sum_{k\geq 0} kz^k=z/(1-z)^2$.

We can now express (\ref{entr}) in the block framework:
$$ H(X^n)=\sum_s N_s (h_s+h^x_s)=n\sum_s p_s(1-p_s) \frac{h^x}{1-p_s} $$
\be =nh^x \sum_s p_s =nh^x. \label{heq}\ee
Finally, the source entropy $H(X^n)=n\sum_s p_s \ln(1/p_s)$ can be alternatively viewed as the sum of entropy
of symbol order ($h_s$) and of block lengths ($h^x_s$).

\subsubsection{InDel Channel with a Single Extrusion}

Assuming the sticky insertion-deletion channel (indel) described above, the sequence of
symbols $s_i$ is unmodified: $Y^n=s_1^{l_1}\ldots {s_N}^{l_N}$. Only the block lengths
are changed in accordance with the noise model ($k_i \to l_i$). Analogously, as in the
previous section, we can determine the entropy of joint distribution $H(X^n,Y^n)$ as:
\be H(X^n)+H(Y^n|X^n)=H(X^n,Y^n)=\sum_s N_s (h_s+h^{xy}_s) \ee
where $h^{xy}_s$ is entropy of pair lengths for input ($k$) and output ($l$) type $s$ blocks:
$$ h^{xy}_s=-\sum_{k,l\geq 1} P_{sk} q_{kl} \ln\left(P_{sk} q_{kl}\right) $$
$$ =h^x_s-\sum_{k\geq 1} P_{sk} \sum_{l\geq 1} q_{kl}\ln(q_{kl}) $$
\be =h^x_s+\sum_{k\geq 1} P_{sk} h^q_k \ee
and $h^q_k:=-\sum_{l\geq 1} q_{kl}\ln(q_{kl})$.

The entropy of output sequence $Y$ and mutual information $I(X^n;Y^n)=H(X^n)+H(Y^n)-H(X^n,Y^n)$ are given by:
$$ H(Y^n)=\sum_s N_s (h_s+h^{y}_s) \qquad $$
\be \textrm{for}\ h^y_s=-\sum_{l \geq 1} \left(\sum_{k\geq 1} P_{sk} q_{kl}\right) \ln\left(\sum_{k\geq 1} P_{sk} q_{kl}\right) \ee
$$ I(X^n;Y^n)=$$
$$=\sum_s N_s \left((h_s+h^x_s)+(h_s+h^y_s)-(h_s+h^{xy}_s)\right) $$
\be =\sum_s N_s (h_s+h^x_s +h^y_s-h^{xy}_s) \ee
$$ H(X^n|Y^n)=H(X^n)-I(X^n;Y^n) $$
$$ =\sum_s N_s(h^{xy}_s-h^y_s) $$
\be =n\sum_s p_s (1-p_s) (h^{xy}_s-h^y_s). \label{exact1} \ee

Asymptotically, the number of typical corrections is given by $\exp(H(X^n|Y^n))$, which grows
exponentially in the length of the sequence $n$. This directly implies the exponentially
decreasing probability of accurate reconstruction. We now discuss how the number of
typical corrections can be reduced (approaching 1) by reading the input sequence multiple
times ($c$). The goal of this analysis is to estimate the number of extrusions we should
perform for unique reconstruction.

\subsubsection{InDel Channel with Multiple Extrusions}

We consider the case of $c$ replicas on each block:
$(l^j_i)_{i=1..N}$ for $j=1,\ldots c$. In this case, we have:
$$h^{xyc}_s=-\sum_{k,l^1,..,l^c\geq 1} P_{sk} q_{kl^1}..q_{kl^c} \ln\left(P_{sk} q_{kl^1}..q_{kl^c}\right) $$
$$ = h^x_s+c\sum_{k\geq 1} P_{sk} h^q_k$$
\noindent
$h^{yc}_s= -\sum_{l^1..l^c \geq 1} \left(\sum_{k\geq 1} P_{sk} q_{kl^1}\cdot..\cdot q_{kl^c}\right)\times$
$$  \times\ln\left(\sum_{k\geq 1} P_{sk} q_{kl^1}\cdot..\cdot q_{kl^c}\right)$$
\noindent
$h^{xyc}_s-h^{yc}_s= \sum_{k\geq 1} P_{sk}\times $
$$ \times\sum_{l^1..l^c \geq 1} q_{kl^1}..q_{kl^c}\ln\left(1+\frac{\sum_{k'\neq k}P_{sk'} q_{k'l^1}..q_{k'l^c}}{P_{sk} q_{kl^1}..q_{kl^c}}\right).$$
Grouping $q_{ki}$ corresponding to the same $i$ by assigning $\tilde{l}^i=\#\{j:l^j=i\}$, using $n!\approx (n/e)^n$, the distribution becomes ($\sum_i \tilde{l}^i=c$):
$$ \sum_{l^1..l^c \geq 1}q_{kl^1}..q_{kl^c}=\sum_{\tilde{l}^1, \tilde{l}^2,...} {c \choose \tilde{l}^1,\tilde{l}^2,...} \prod_{i\geq 1} q_{ki}^{\tilde{l}^i} $$
$$ \approx
\sum_{\tilde{l}^1,..}
\prod_{i\geq 1} \left(\frac{c}{\tilde{l}^i}\right)^{\tilde{l}^i} q_{ki}^{\tilde{l}_i} $$
$$ = \sum_{\tilde{l}^1,..}\exp\left(-c\sum_{i\geq 1} \frac{\tilde{l}^i}{c} \ln \left(\frac{\tilde{l}^i/c}{q_{ki}}\right)\right).
$$
The sum in bracket is the Kullback-Leibler (asymmetric) divergence: $D_{KL}\left(\left\{\tilde{l}^i/c\right\}_i\ ||\ \left\{q_{ki}\right\}_i\right)$ - the exponent is asymptotically (large $c$) dominated by $\tilde{l}^i/c=q_{ki}$ distribution. To find an
approximation of the formula $h^{xyc}_s-h^{yc}_s$, we focus only on these distributions:
$$h^{xyc}_s-h^{yc}_s\approx$$
 $$\approx \sum_{k\geq 1} P_{sk}\ln\left(1+\frac{\sum_{k'\neq k}P_{sk'} \left(\prod_{l\geq 1} q_{k'l}^{q_{kl}}\right)^c}{P_{sk}\left(\prod_{l\geq 1} q_{kl}^{q_{kl}}\right)^c}\right) $$
$$=\sum_{k\geq 1} P_{sk}\ln\left(1+\sum_{k'\neq k} \frac{P_{sk'}}{P_{sk}} e^{-c\cdot d_{kk'}}\right)$$
 $$\approx \sum_{k\geq 1} \sum_{k'\neq k}P_{sk'} e^{-c\cdot d_{kk'}}$$
where $d_{kk'}$ is the Kullback-Leibler  divergence:
$$d_{kk'}:=-\ln\left(\prod_{l\geq 1} \left(\frac{q_{k'l}}{q_{kl}}\right)^{q_{kl}}\right)$$
$$ =\sum_{l\geq 1} q_{kl}\ln\left(\frac{q_{kl}}{q_{k'l}}\right)
 =D_{KL}\left(\left\{q_{kl}\right\}_l \ ||\ \left\{q_{k'l}\right\}_l \right)$$
Since $H(X|Y)=H(X,Y)-H(Y)$ and $P_{sk}=p_s^{k-1}(1-p_s)$, we can write:
$$ H(X^n|Y^{nc})=n\sum_s p_s (1-p_s) (h^{xyc}_s-h^{yc}_s) $$
\be \approx n\sum_s (1-p_s)^2 \sum_{k\geq 1} \sum_{k'\neq k}p_s^{k'} \exp(-c\cdot d_{kk'}) \label{con1}\ee
\be \approx n\exp(-c\cdot d^{min})\cdot \sum_s (1-p_s)^2p_s^{k_0'}\label{con2}\ee
where $d^{min}:=\min_{k'\neq k} d_{kk'}=d_{k_0 k'_0}$ is the minimal value of $d$ (if the minimum exists).

The distance $d_{kk'}$ describes the similarity between the results of reading blocks
having original lengths $k$ and $k'$. It quantifies the likelihood of mistakenly
identifying a $k$ length block as a $k'$ length block. The smaller it is, the faster
is the growth of number of typical corrections $(\exp(H(X^n|Y^{nc})))$ with $k$ erroneously
replaced by $k'$. The smallest distance $(d^{min})$ corresponds to the most likely
mistake, and it asymptotically dominates the growth of typical corrections.

The use of multiple extrusions $(c)$ allows us to reduce the exponent in
the number of typical corrections $\exp(H(X^n|Y^{nc}))$. For unique reconstruction,
the number of typical corrections must approach 1. Consequently, we choose $c$ such
that $n\cdot \exp(-c\cdot d^{min})$ is of order of 1. From this, we see that
the number of extrusions should grow logarithmically in sequence length:
$c\approx \ln(n)/d^{min}$. This important result establishes the feasibility
of low-overhead sequencing using nanopore sequencers.

\paragraph{Analysis of the substitution-only case} \label{substitution}
The framework presented above for indel errors can also be extended to substitution-only
errors: a reference base $s\in\mathcal{A}$ has probability $q_{s\ell}$ to be read as
base $\ell\in\mathcal{A}$, where $|\mathcal{A}|=4$ is the set of 4 bases. The expected number of occurrences of
base $s$ in a sequence of length $n$ is given by $\tilde{N}_s=np_s$. We analogously have:
$$h^{xyc}_s-h^{yc}_s=$$
$$= \sum_{\ell^1..\ell^c \geq 1} q_{s\ell^1}..q_{s\ell^c}\ln\left(1+\frac{\sum_{s'\neq s}q_{s'\ell^1}..q_{s'\ell^c}}{q_{s\ell^1}..q_{s\ell^c}}\right)$$
$$\approx \ln\left(1+\sum_{s'\neq s}\left(\prod_{\ell\in\mathcal{A}} \left(\frac{q_{s'\ell}}{q_{s\ell}}\right)^{q_{s\ell}}\right)^c\right)$$

$$ H(X^n|Y^{nc})=\sum_s \tilde{N}_s (h^{xyc}_s - h^{yc}_s)\approx $$
\be \approx n\sum_s p_s \sum_{s'\neq s} \exp(-c\cdot d_{ss'}). \ee
where $d_{ss'}=D_{KL}(\{q_{s\ell}\}_\ell||\{q_{s'\ell}\}_\ell)$.

Analogously to the indel case, $\exp(H(X^n|Y^{nc}))$ is the size of distortion ball - the
number of typical reconstructions. For unique reconstruction, we need this size to be of
order one. If there exists $d^{min}=\min_{s\neq s'} d_{ss'}$,
the number of reads required for unique reconstruction is approximately
$c\approx \ln(n)/d^{min}$; i.e., logarithmic in sequence length.

\section{Experimental Results} \label{experiment}

We present a simulation study of the implications of our analysis on real-world sequencing
experiments. We consider three models for our channel -- the first model is a sticky channel with
exponential distribution and the second, an independent insertion-deletion channel.
Finally, we also use these general considerations for substitution-only case.
In each case, we examine the bound on length of sequence for accurate reconstruction, the number
of replicas needed for larger reconstructions, and the Kullback-Leibler divergences. The goal
of these studies is to demonstrate that for a wide class of channel characteristics: (i) the
length of sequence that can be accurately reconstructed in single read is small; (ii) the
number of required replicas for longer reconstructions is a slowly growing function
(logarithmic); and (iii) even in the presence of high indel error rates, nanopore
sequencers can accurately reconstruct sequences with required number of replicas.

We consider a binary equi-probable input -- $m=2$, $p_0=p_1=1/2$. The behavior of conditional
entropy $H(X|Y)$ is primarily determined by the $q_{ks}$ distribution. The
results in this section can be naturally generalized to any alphabet and probability
distribution.

\subsection{Exponential Insertion-Deletion Error Model}
%


We will first consider a sticky channel with exponential distribution for the error
probabilities -- for some $0<q<1$: $$q_{kl}=q^{|k-l|} \frac{1-q}{1+q-q^k}$$
The second term in the product is for normalizing the probability.

\begin{figure*}
    \centering
        \includegraphics[width=10cm]{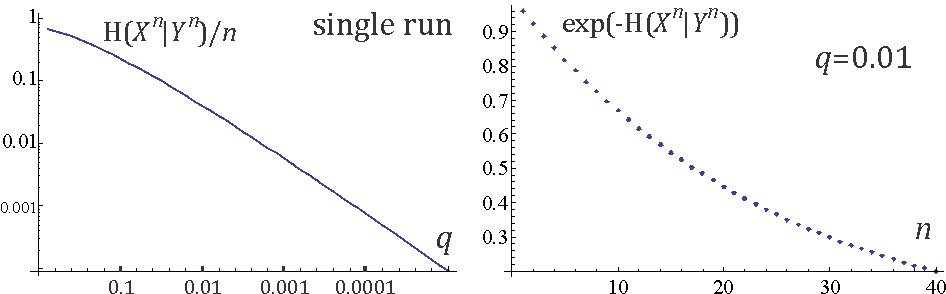}\
        \caption{Left Panel: Conditional entropy for single extrusion between input
sequence (input to the nanopore sequencer) and the output sequence (observed sequence).
This is derived from Equation (\ref{exact1}). Right Panel: The probability that we
select the correct typical correction for $q=0.01$ and increasing value of $n$.}
        \label{sr}
\end{figure*}

We first consider the single run case. Equation (\ref{exact1}) allows us to calculate
conditional entropy $H(X^n|Y^n)$, describing the growth in the number of typical corrections:
$\exp(H(X^n|Y^n))$. The left Panel of Figure \ref{sr} presents values of conditional
entropy for various values of $q$. Assuming correction procedure as taking a random
typical correction, the Right Panel of this figure presents the probability of obtaining
the right correction. This probability drops exponentially with the length of the
sequence, making a single run approach impractical for longer sequences.

This limitation can be handled by performing multiple extrusions of the same sequence.
We use Equation (\ref{con1}) to find conditional entropy in this case. This requires
finding Kullback-Leibler divergences between $\{q_{kl}\}_l$ distributions for different
original block lengths $k$. Table~1  presents some of these values for $q=0.5$.

\begin{table}
\begin{center}
{\footnotesize
\begin{tabular}{c|ccccccc}
  $k' \rightarrow$ & $1$ & $2$ & $3$ & $4$ & $5$ & $6$ & $7$\\
  $k \downarrow$ & & & & & & & \\ \hline
  $1$ & 0 & 0.223 & 0.665 & 1.123 & 1.857 & 2.512 & 3.194 \\
  $2$& \textbf{0.193} & 0 & 0.234 & 0.694 & 1.270 & 1.905 & 2.568  \\
  $3$ & 0.567 & 0.220 & 0 & 0.233 & 0.696 & 1.274 & 1.909 \\
  $4$ & 1.054 & 0.644 & 0.227 & 0 & 0.232 & 0.695 & 1.273\\
  $5$ & 1.621 & 1.181 & 0.671 & 0.229 & 0 & 0.232 & 0.694 \\
  \end{tabular}
}
\end{center}
\caption{ Kullback-Leibler divergence for $q=0.5$, exponential distribution, and various
original block lengths: $k$, $k'$.
The smallest value allows to conclude that we asymptotically need
$c\approx \ln(n)/0.193$ reads for unique reconstruction. }
\label{kl_distance_1}
\end{table}

\begin{figure*}
    \centering
        \includegraphics[width=10cm]{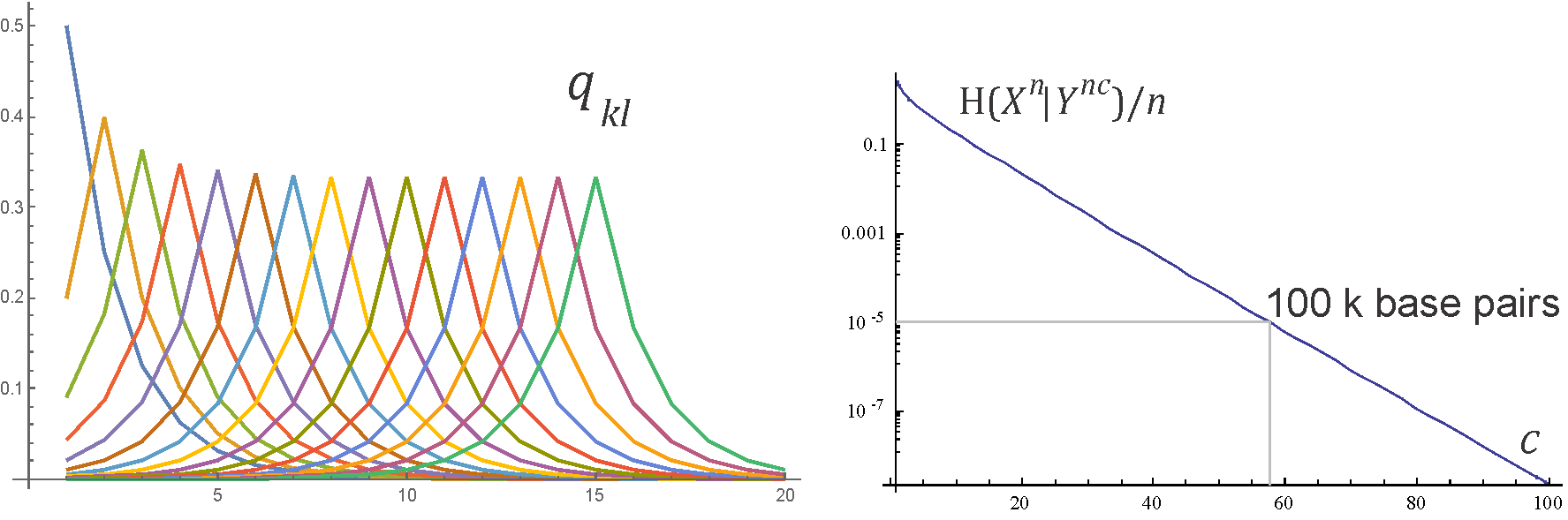}\
        \caption{Left Panel: $\{q_{kl}\}_l$ distributions for $q=0.5$ and $k=1$ to 15.
        Right Panel: (\ref{con1}) approximation for $q=0.5$ and different numbers of copies $c$.
        For example for $n=10^5$ length sequence we need less than $60$ reads for unique reconstruction.
        }
        \label{abs}
\end{figure*}

The minimal distance is $d^{min}=d_{21}\approx 0.193$, and corresponds to
misinterpreting original $k=2$ sequence as $k'=1$. This intuitively stronger
overlap of the first two distribution can be observed in the Left Panel of
Figure \ref{abs}, containing $\{q_{kl}\}_l$ for the first 15 values of $k$.
The distance between farther neighboring distributions is nearly the same; i.e.,
misreading a block of length five nucleotides as six, is as likely as an input block
of 100 nucleotides being read as 101.

The right panel of Figure~\ref{abs} shows conditional entropy as a function of
number of replicas $c$, for $q=0.5$. There are important observations drawn from
this figure: (i) the nearly linear nature of the curve shows that the number of
replicas is almost logarithmic in the read length ($H(X^n|Y^n)/n$ asymptotically
behaves as $\exp(-c\cdot d^{min}$); and (ii) for realistic reconstruction
lengths (say, 100K bases), the number of replicas is relatively small (less than 60).
These are important results that establish the feasibility of nanopore sequencers
for accurate low-cost construction of long reads.

\begin{figure*}
    \centering
        \includegraphics[width=10cm]{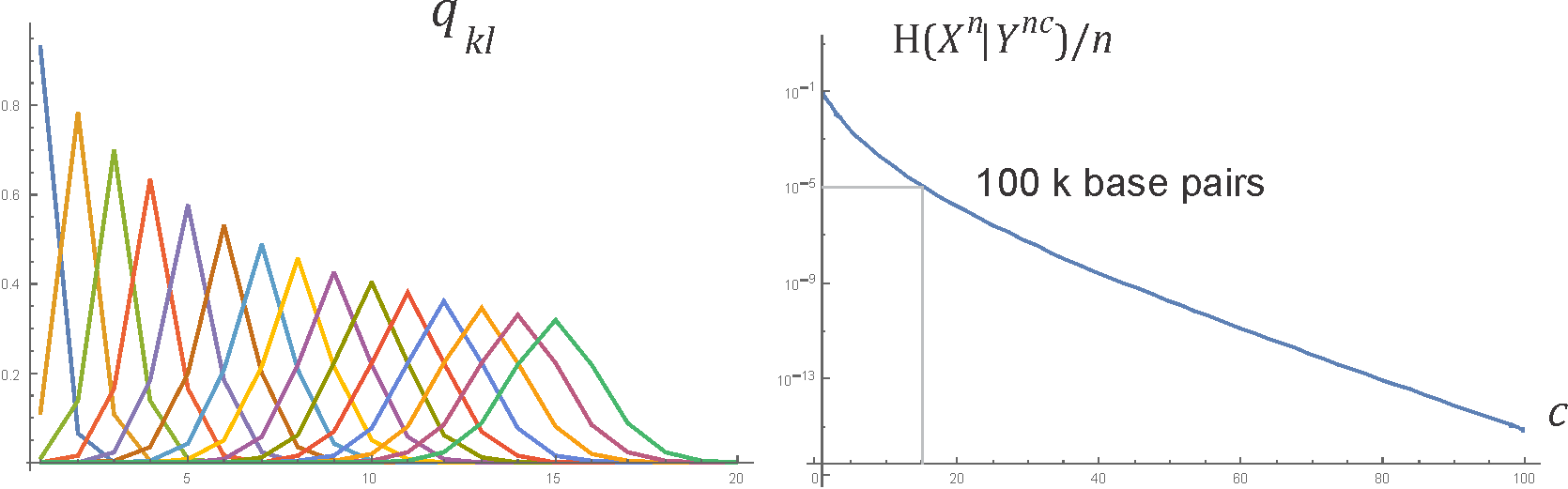}\
        \caption{Left Panel: The first 15 $q$ distributions for $\epsilon=0.06$ for the
independent insertion-deletion channel. Right Panel: Approximation of joint entropy for
$\epsilon=0.1$ and different numbers of copies $c$ (from Equation (\ref{con1}).
 For example for $n=10^5$ length sequence we need less than $20$ reads for unique reconstruction.}
        \label{ans}
\end{figure*}

\subsection{Independent Insertion-Deletion Channel Model} To demonstrate the robustness of
our results we now consider a different channel model -- the independent insertion-deletion
channel. In this model, for each nucleotide, there is probability $\epsilon>0$, that the
symbol is deleted. There is also an identical probability that the symbol is duplicated.
It follows that there is a probability $1-2\epsilon$ that the symbol is sequenced without
an error. For this model, $q_{kl}$, for given $k$, is the convolution of $k$ such random
variables, additionally truncated to enforce that it is a sticky channel; i.e., $q_{k0}=0$.
For large $k$, $q_{kl}$ approaches the Gaussian distribution with standard deviation
$\sqrt{2\epsilon k}$.

Figure~\ref{ans} shows the first 15 distributions ($q_{kl}$) for this model. It is illustrative
to note that unlike many previous models in which error rates are invariant on block length,
larger block lengths have higher insertion-deletion error rates in our model. Table~2 presents
the approximated first values of similarities of $q_{kl}$ for different values of $k$, for
$\epsilon=0.06$, derived from Jain et al.~\cite{jain}. The infinite values in
the table correspond to difference in support (one of values is zero).
For example, the value $k=2$ can lead to blocks of length 1, 2, 3, or 4, while
$k=1$ can lead only to blocks of length 1 or 2. Many reads of length $k=2$ will lead to
some blocks of length 3 or 4. These are in addition to original blocks of length $k=1$, which
may be misinterpreted in the assumed model.
We note that the distributions become closer to their own neighbors as $k$ grows:
$\lim_{k\to\infty} d_{k k+1}=0$, the minimal nonzero distance $d^{min}$ does not exist.
In other words, distinguishing between $k$ and $k+1$ becomes more difficult with
increasing $k$, and their difference vanishes asymptotically.
Consequently, as we can observe from the right panel of Figure~\ref{ans}, the required
number of replicas grows slower than logarithm of sequence length.
Figure~\ref{ans} shows the conditional entropy of the input and output sequences for different
numbers of replicas ($c$). long sequences (100K nucleotides), we need even fewer replicas
(less than 20, in this case, compared to 60 for the exponential model).

\begin{table}
\begin{center}
\begin{tabular}{c|ccccccc}
  $k' \rightarrow$ & $1$ & $2$ & $3$ & $4$ & $5$ & $6$ & $7$\\
  $k \downarrow$ & & & & & & & \\ \hline
  $1$ & 0 & 1.879 & 4.247 & 6.747 & 9.319 & 11.94 & 14.58 \\
  $2$ & $\infty$ & 0 & 1.394 & 3.473 & 5.760 & 8.160 & 10.63 \\
  $3$ & $\infty$ & $\infty$ & 0 & 1.091 & 2.931 & 5.032 & 7.295\\
  $4$ & $\infty$ & $\infty$ & $\infty$ & 0 & 0.886 & 2.219 & 4.024\\
  $5$ & $\infty$ & $\infty$ & $\infty$ & $\infty$ & 0 & 0.739 & 2.220\\
\end{tabular}
\end{center}
\label{similarities}
\caption{Similarity of $q_{kl}$ for different values of $k$ for the independent
insertion-deletion channel.
Infinities denotes that asymptotic empirical probability e.g.
for $k=2$ cannot be misinterpreted as $k=1$.  }
\end{table}

\subsection{Substitution-only case}
\begin{figure}
    \centering
        \includegraphics[width=8cm]{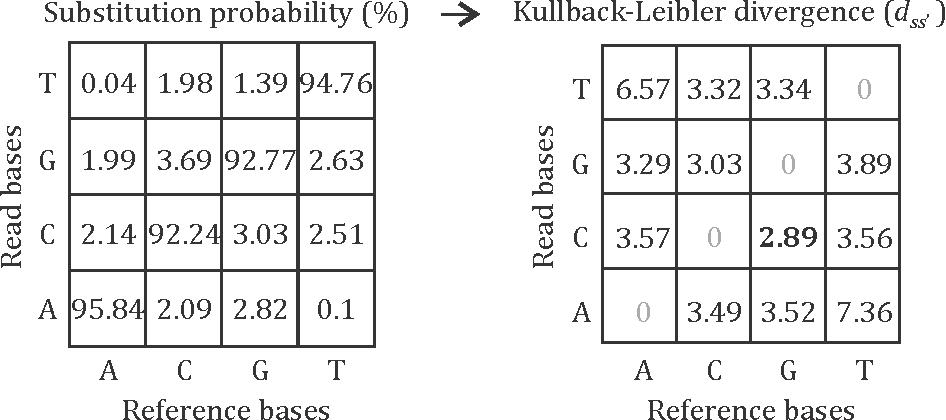}\
        \caption{Left Panel: Probabilities of substitutions taken from~\cite{jain}. Right Panel: Kullback-Leibler divergence between these probability distributions. Asymptotically, the case that base G will be wrongly read as base C will dominate. The required number of reads for unique reconstruction of substitution-only case grows like $\ln(n)/2.89$.
         }
        \label{subs}
\end{figure}

We finally use our approach to the substitution-only case from Section \ref{substitution}.
Figure \ref{subs} contains probabilities of substitution from~\cite{jain}.
In this case, the required number of reads is given by:
$$c\approx n\sum_s p_s \sum_{s'\neq s} \exp(-c\cdot d_{kk'})$$
what is $c\approx ln(n)/2.89$ in the assumed case.

\section{Related Research}

Technologies underlying nanopore sequencers have been investigated for over
a decade~\cite{deamer, butler}. Commercial platforms based on these
technologies have only recently been announced -- with Oxford Nano being
the leading platform. An excellent introduction to this platform
is available at:
\url{https://www.nanoporetech.com/technology/analytes-and-applications-dna-rna-proteins/dna-an-introduction-to-nanopore-sequencing}. There have been preliminary efforts aimed
at characterizing the performance of nanopore sequencing platforms in terms
of error rate, error classification, and run lengths~\cite{e1,Odonnell,e3,e4}. A
consensus emerges from these studies that the primary error mode in
nanopore sequencers is deletion errors and that the error rate is approximately
4\% with a read length of over 150K bases. These studies provide important
data that is used to build our insertion-deletion channel.

Churchill and Waterman~\cite{churchill} address similar questions
to ours for substitution channels. A key differentiating aspect of their work is that
they reconstruct a sequence of read values for a given position as a consensus; i.e.,
the most frequent value across replicates. In contrast, we focus of the most probable
reconstruction. This leads to comparison of obtained empirical probability distribution
of reads for given value, with probability distributions expected for different
reference values (using cross entropy, equation (\ref{cross})). The final formulae for reconstruction bounds
contain Kullback-Leibler divergence between probability distributions for different
reference values quantitatively describing the difficulty of distinguishing them.

Error characteristics and models for nanopore sequencers have been
recently studied by O'Donnell et al~\cite{Odonnell}. In this study, the authors
investigate error characteristics, and build a statistical model
for errors. They use this model to show, through a simulation study,
that replicated extrusion can be used to improve error characteristics.
In particular, they show that using their model, it is possible to achieve
99.99\% accuracy by replicating the read 140 times. This empirical study provides
excellent context for our analytical study, which provides rigorous
bounds and required replication rates.

There have been a number of efforts aimed at analyzing the error characteristics
of current generation of sequencing technologies, including the 454 and
PacBio sequencers~\cite{koren2012hybrid, carneiro2012pacific, ono2013pbsim}.
These efforts are primarily aimed at the problem of alignment of relatively short
fragments - of localizing their positions in a longer sequence. In contrast,
due to longer reads obtained by nanopore sequencing, in this paper, we
do not deal with the issue of localizing positions of the fragments and
focus on the problem of correcting indel and substitution errors. These
considerations can be also used for improving accuracy from multiple
(aligned) reads of any sequencing technique. Our approach takes an information
theoretic view to the problem. In doing so, we are able to establish fundamental
bounds on the performance envelope of the modeled nanopore sequencer. To the
best of our knowledge, this paper represents the first information theoretic
formulation of its kind.

There has been significant work on different channels, their capacities,
and error characteristics since the work of
Shannon. Of particular relevance to our results is the work in deletion
channels~\cite{DBLP:journals/corr/abs-1102-0040}. As mentioned, the
capacity of independent deletion channels is as-yet unknown. There have
been efforts aimed at error correction in insertion-deletion channels in
the context of communication, storage, and RFID systems~\cite{guang12ita}.

\section{Discussion and Conclusion}
\label{discussion}

In this paper, we present a novel modeling methodology based on the abstraction
of a nanopore sequencer as an information theoretic channel. We use our methodology
to show a number of important results: (i) the indel and substitution error rates
of the nanopore sequencer limit the sequence length that can be accurately reconstructed;
(ii) replicated extrusion through the nanopore is an effective technique for
increasing the accurate reconstruction length; (iii) the
number of replicas is a slow function of the sequence length (logarithmic in
sequence length for both indel and substitution errors), enabling nanopore sequencers
to accurately reconstruct long sequences.

We demonstrate our results for a wide class of error models and show that our
analyses is robust. We note that our analyses accounts for indel and substitution
errors separately. In this sense, it provides lower bounds for a model that
integrates both of these error modes into a single model (channel).

\section*{Acknowledgments}

This work is partially supported by the NSF Science and Technology Center for
Science of Information Grant CCF-0939370, NSF Grants DMS-0800568, DBI-0835677, IOS-1124962,
and CCF-0830140, and MNSW grant DEC-2013/09/B/ST6/02258.

\vspace*{-0.1in}
\bibliographystyle{plain}
\bibliography{cites}

\begin{thebibliography}{10}

\bibitem{butler}
T.~Butler, M.~Pavlenok, I.~Derrington, M.~Niederweis, and J.~Gundlach.
\newblock Single-molecule dna detection with an engineered mspa protein
  nanopore.
\newblock {\em Proceedings of the National Academy of Science},
  105(52):20647--20652, 2008.

\bibitem{carneiro2012pacific}
Mauricio~O Carneiro, Carsten Russ, Michael~G Ross, Stacey~B Gabriel, Chad
  Nusbaum, and Mark~A DePristo.
\newblock Pacific biosciences sequencing technology for genotyping and
  variation discovery in human data.
\newblock {\em BMC genomics}, 13(1):375, 2012.

\bibitem{churchill}
Gary~A Churchill and Michael~S Waterman.
\newblock The accuracy of dna sequences: estimating sequence quality.
\newblock {\em Genomics}, 14(1):89--98, 1992.

\bibitem{deamer}
D.~Deamer and D.~Branton.
\newblock Characterization of nucleic acids by nanopore analysis.
\newblock {\em Acc Chem Res}, 35(10):817--825, 2002.

\bibitem{mitz}
E.~Drinea and M.~Mitzenmacher.
\newblock Improved lower bounds for the capacity of i.i.d. deletion and
  duplication channels.
\newblock {\em {IEEE Transactions on Information Theory}}, 53:8:2693--2714,
  2007.

\bibitem{e3}
E.~Hayden.
\newblock Nanopore genome sequencer makes its debut.
\newblock {\em Nature News}, Feb. 2012.

\bibitem{jain}
Miten Jain, Ian~T Fiddes, Karen~H Miga, Hugh~E Olsen, Benedict Paten, and Mark
  Akeson.
\newblock Improved data analysis for the minion nanopore sequencer.
\newblock {\em Nature methods}, 2015.

\bibitem{DBLP:journals/corr/abs-1102-0040}
Ian~A. Kash, Michael Mitzenmacher, Justin Thaler, and Jonathan Ullman.
\newblock On the zero-error capacity threshold for deletion channels.
\newblock {\em CoRR}, abs/1102.0040, 2011.

\bibitem{koren2012hybrid}
Sergey Koren, Michael~C Schatz, Brian~P Walenz, Jeffrey Martin, Jason~T Howard,
  Ganeshkumar Ganapathy, Zhong Wang, David~A Rasko, W~Richard McCombie, Erich~D
  Jarvis, et~al.
\newblock Hybrid error correction and de novo assembly of single-molecule
  sequencing reads.
\newblock {\em Nature biotechnology}, 30(7):693--700, 2012.

\bibitem{Lanctot:2000:EDS:338219.338586}
J.~Kevin Lanctot, Ming Li, and En-hui Yang.
\newblock Estimating dna sequence entropy.
\newblock In {\em Proceedings of the Eleventh Annual ACM-SIAM Symposium on
  Discrete Algorithms}, SODA '00, pp. 409--418, Philadelphia, PA, USA, 2000.
  San Francisco, California, USA, Society for Industrial and Applied
  Mathematics.

\bibitem{e1}
A.~Mikheyev and M.~Tin.
\newblock A first look at the oxford nanopore minion sequencer.
\newblock {\em Molecular Ecology Resources}, 14(6):1097--1102, Nov. 2014.

\bibitem{Odonnell}
C.~O'Donnell, H.~Wang, and W.~Dunbar.
\newblock Error analysis of idealized nanopore sequencing.
\newblock {\em Electrophoresis}, 34(15):2137-44, 2013.

\bibitem{ono2013pbsim}
Yukiteru Ono, Kiyoshi Asai, and Michiaki Hamada.
\newblock Pbsim: Pacbio reads simulator—toward accurate genome assembly.
\newblock {\em Bioinformatics}, 29(1):119--121, 2013.

\bibitem{sequencing_survey}
M.~Quail, M.~Smith, P.~Coupland, T.~Otto, S.~Harris, T.~Connor, A.~Bertoni,
  H.~Swerdlow, and Y.~Gu.
\newblock A tale of three next generation sequencing platforms: comparison of
  ion torrent, pacific biosciences and illumina miseq sequencers.
\newblock {\em BMC Genomics}, 13:341, 2012.

\bibitem{e4}
J.~Schreiber, Z.~Wescoe, R.~Abu-Shumays, J.~Vivian, B.~Baatar, K.~Karplus, and
  Mark Akeson.
\newblock Error rates for nanopore discrimination among cytosine,
  methylcytosine, and hydroxymethylcytosine along individual dna strands.
\newblock {\em Proceedings of the National Academy of Science}, 110(47), Nov.
  19, 2013.

\bibitem{guang12ita}
Guang Yang, Angela~I. Barbero, Eirik Rosnes, and Yvind Ytrehus.
\newblock Error correction on an insertion/deletion channel applying codes from
  rfid standards.
\newblock {\em {ITA}}, 2012.

\end{thebibliography}

\end{document}